\documentclass[aip,jcp,reprint]{revtex4-1}

\usepackage{graphicx}
\usepackage{amsmath}
\usepackage{amssymb}

\begin{document}

\title{A computationally efficacious free-energy functional\\
for studies of inhomogeneous liquid water}

\author{Ravishankar Sundararaman} \email{rs596@cornell.edu}
\author{Kendra Letchworth-Weaver}
\author{T. A. Arias}
\affiliation{Cornell University Department of Physics}
\date{\today}

\begin{abstract}
We present an accurate equation of state for water
based on a simple microscopic Hamiltonian,
with {\em only} four parameters that are
well-constrained by bulk experimental data.
With one additional parameter for the range of interaction,
this model yields a computationally efficient
free-energy functional for inhomogeneous water
which captures short-ranged correlations,
cavitation energies and, with suitable long-range corrections,
the non-linear dielectric response of water,
making it an excellent candidate for studies of
mesoscale water and for use in {\em ab initio} solvation methods.
\end{abstract}

\maketitle

\section{Introduction}
The emergence of several macroscopic phases of water
with distinct microscopic structures \cite{IceReview,AmorphousIceReview}
from relatively simple molecular interactions
places this liquid at the forefront of interesting
unsolved problems in the study of condensed matter.
The structure of water around microscopic objects
differs significantly from the bulk,\cite{ConfinedWater}
and such effects play a critical role in the structure
of proteins \cite{WaterProteinFolding} and in chemical
reactions at catalyst surfaces.\cite{WaterCatalyst}

Current computational approaches to systems sensitive to
liquid structure most often employ molecular dynamics simulations.
{\em Ab initio} molecular dynamics,\cite{CPMD} which treats all the valence
electrons in the system quantum mechanically, is relatively accurate
but prohibitively expensive for all but the smallest of systems.
Hybrid approaches that combine electronic structure methods
for part of the system with classical molecular dynamics
simulations for the fluid can handle larger systems,
but require empirical models for both the electron-fluid coupling
and the classical pair potentials for the fluid.
These molecular dynamics methods are inherently expensive
due to the sampling required for thermodynamic averages which,
when coupled to an electronic system,
necessitates repeated electronic structure calculations.
In addition, the thermodynamic integration required to calculate
free energies, which are necessary for analyzing chemical reaction pathways,
significantly exacerbates the cost of such methods.

Efficient theories for the equilibrium properties of liquids,
on the other hand, deal directly with average densities
instead of individual molecular configurations.
Integral equation theories give accurate structures of
inhomogeneous fluids,\cite{IntEqnInhomogeneous} but still prove
relatively expensive, particularly for estimating free energies.
The most direct approach to free energies is classical
density-functional theory, a method based on approximations to the
exact free-energy functional of the liquid,\cite{MerminTheorem}
which has the added advantage of being readily coupled
to electronic density-functional theory within the
framework of joint density-functional theory.\cite{JDFT}
The most accurate, currently available functionals for polar molecular fluids
such as water,\cite{WDA,WaterFreezingDFT,LischnerHCl,LischnerH2O,WuIntegralEqnCDFT}
however, rely on direct correlations (from neutron scattering or computer simulation)
at each temperature and pressure of interest,
restricting their efficiency and applicability.

This work addresses the need for a computationally efficacious
microscopic theory of water that is capable of providing
accurate free energies under inhomogeneous conditions
without the dependence on fluid structure data.
The strategy is to identify a simple {\em effective microscopic Hamiltonian}
which (a) reproduces the equation of state for homogeneous water and
(b) is readily represented by a free-energy functional even in the inhomogeneous case.

Statistical Associating Fluid Theory,\cite{SAFT-DFT}
based on Wertheim's thermodynamic pertubation theory,\cite{WertheimTPT}
is one such approach which has been successfully applied to the study of
vapor-liquid interfaces,\cite{SAFT-Tension} with model parameters
for water determined from the equation of state.\cite{SAFT-Water}
However, the predictions of this theory for the inhomogeneous fluid
have not yet included quantities of interest in solvation methods
such as pair correlations, cavitation energies and dielectric response,
partly due to the relative complexity of the model Hamiltonian.
Below, we develop an alternate simpler Hamiltonian based upon
microscopic intuition about hydrogen bonding, and we demonstrate that
the resulting functional (also based on Wertheim theory) leads to a 
relatively accurate free-energy description of inhomogeneous water,
especially given the simplicity of the underlying model.

\section{Model molecular Hamiltonian and the equation of state for water}
Within the constraints of condition (b) above, a natural starting point
would be the standard approach of perturbation about the hard-sphere fluid,
for which Fundamental Measure Theory\cite{RosenfeldFMT,TarazonaFMT,WhiteBearFMT_MarkII,FMTreview}
provides a highly accurate functional. The hard-sphere diameter required 
to reproduce bulk properties can be inferred from the excluded volume
in the equation of state, and fits\cite{JeffAustinEOS}
to experimental data suggest a value that strongly decreases 
with temperature and is $\sim3.3$\AA\ at $298$~K.
This is clearly incompatible with the almost temperature-independent
$\sim2.8$\AA\ location of the first peak in the experimentally observed 
oxygen-oxygen radial distribution.\cite{SoperEPSR}

\begin{figure}
\includegraphics[width=1.375in]{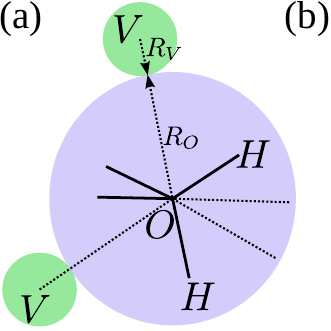}\includegraphics[width=2in]{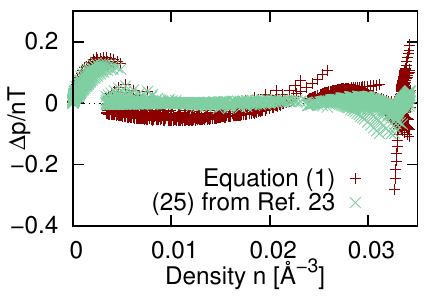}
\caption{(a) Tangentially bonded hard-sphere model for liquid water:
$O$ (oxygen) sphere with two $V$ spheres in contact
diametrically opposite to the  $H$ (hydrogen) sites
(b) Residual for the equation of state (\ref{eqn:EOS})
fit to experimental data,\cite{BulkData,SupercooledData}
compared to the semi-empirical Jeffery-Austin
equation of state.\cite{JeffAustinEOS} \label{fig:trimer}\label{fig:eosFit}}
\end{figure}

This incompatibility stems from the discrepancy between the
close-packed coordination of the hard-sphere fluid
and the tetrahedral coordination favored by water.
Water prefers the formation of open tetrahedral networks
at lower temperatures, which leads to empty space, ``voids'',
within cages of water molecules, as manifested by the 
temperature-dependent excess excluded volume in the equation of state.
Consequently, we propose a reference fluid consisting of a compound object
(FIG.~\ref{fig:trimer}(a)): a hard sphere of radius $R_O$ at the
$O$ (oxygen) site with smaller spheres of radii $R_V$ at two void 
sites $V$ placed in contact (at a distance $\sigma_{OV}=R_O+R_V$)
along two of the conjugate tetrahedral directions to the hydrogen bond directions.
For our model, we take the $O$-$H$ distance to be 1~\AA\ with a tetrahedral $H$-$O$-$H$ angle,
as in the frequently employed SPC/E interatomic potential model.\cite{SPCE}
The geometry of this compound object is chosen to encourage
closest approach along the hydrogen bond directions.

Our ansatz for the intermolecular Hamiltonian is the repulsive pair potential
corresponding to the tangentially bonded hard-sphere trimer of FIG.~\ref{fig:trimer}(a), 
perturbed by an isotropic attractive pair potential $U_a(r)$ between the $O$ sites.
The equation of state of this fluid is well approximated by 
\begin{flalign}
p(n,T) &= p_{\textrm{id}} + p^{\textrm{ex}}_{\textrm{HS}}
	- 2 n^2  T \left. \frac{\partial \log \left.
		g^{\textrm{HS}}_{OV}\right|_{\sigma_{OV}}}{\partial n} \right|_T
	- \frac{\kappa n^2}{2}, \label{eqn:EOS}
\end{flalign}
where the first three terms correspond to the bonded hard-sphere equation
of state,\cite{BHS_EOS} and the final term is the mean-field contribution
from the as yet undetermined attractive perturbation $U_a(r)$,
with $\kappa\equiv -\int\textrm{d}r 4\pi r^2 U_a(r)$.

The bonded hard sphere equation of state is based on Wertheim
perturbation theory \cite{WertheimTPT} about the hard sphere mixture,
consisting of density $n$ of $O$-spheres and $2n$ of $V$-spheres.
The pressure of this reference  system is
$p_{\textrm{HS}} = 3nT + p^{\textrm{ex}}_{\textrm{HS}}$,
where we separate and collect the $\mathcal{O}(n)$ ideal gas parts
to elucidate the connection with the density functional
(\ref{eqn:Functional}). For $p^{\textrm{ex}}_{\textrm{HS}}$,
we employ the accurate generalization\cite{HardSphereEOS_CSIII}
of the Carnahan-Starling excess pressure to hard sphere mixtures
\begin{flalign}
p^{\textrm{ex}}_{\textrm{HS}} &=  T \left[ \frac{n_0 n_3}{1-n_3}
		+ \frac{n_1 n_2}{(1-n_3)^2} \left(1+\frac{n_3^2}{3}\right) \right. \nonumber\\
	&+ \left. \frac{n_2^3}{12\pi(1-n_3)^3} \left(1-\frac{2n_3}{3}+\frac{n_3^2}{3}\right) \right], \label{eqn:pExHS}
\end{flalign}
where $n_0=3n$, $n_1=(R_O+2R_V)n$, $n_2=4\pi(R_O^2+2R_V^2)n$ and
$n_3=\frac{4\pi}{3}(R_O^3+2R_V^3)n$ are the uniform fluid fundamental measures.

First order Wertheim perturbation for the bonding constraints accounts
for the fixed $O$-$V$ separation and not the $V$-$O$-$V$ angle;
nonetheless it has been shown to well approximate
the equation of state of objects with this geometry.\cite{BHS_EOS}
We accumulate its contribution at $\mathcal{O}(n)$
into the first term of (\ref{eqn:EOS}): this exactly
corrects the ideal gas mixture value of $3nT$ to the
rigid-molecule ideal gas value of $p_{\textrm{id}}=nT$.
(We use this fact later to restore the bond angle constraints in
the intramolecular geometry of the inhomogeneous fluid.)
The remaining contribution of this perturbation, the third term of
(\ref{eqn:EOS}), corrects the excluded volume effects of the hard
sphere mixture to account for the $O$-$V$ distance constraints.
There,
\begin{flalign}
\left. g^{\textrm{HS}}_{OV}\right|_{\sigma_{OV}}
	= \frac{1}{1-n_3} + \frac{n_2 R_{\textrm{hm}}}{(1-n_3)^2}
	+ \frac{2 (n_2 R_{\textrm{hm}})^2}{9(1-n_3)^3},
\label{eqn:gOV}
\end{flalign}
is the contact value of the $O$-$V$ radial distribution in the
hard sphere mixture with $R_{\textrm{hm}} = R_O R_V / \sigma_{OV}$.

As motivated earlier, the temperature dependence of the exclusion volume
is a critical feature of the equation of state for water.\cite{JeffAustinEOS}
Because the location of the first peak in the $O$-$O$ radial distribution does
not change appreciably with temperature, we attribute this dependence
to changes in the radii of the $V$ spheres, modeled as a decreasing function
$R_V(T) = R_V^{(0)} \textrm{e}^{-T/T_V}$ to qualitatively capture
the effect of the empty spaces in the open tetrahedral network.
This leads to a model equation of state (\ref{eqn:EOS}) with only
four adjustable parameters ($R_V^{(0)}$, $T_V$, $\kappa$, and $R_O$),
which we fit to experimental data for the bulk liquid and vapor \cite{BulkData}
including data for the supercooled liquid.\cite{SupercooledData}

The root mean-square error in the ratio of the pressure to the
ideal-gas pressure, $p/nT$, is $4.8\times10^{-2}$ for the current 4-parameter fit,
which compares very favorably with $2.9\times10^{-2}$ for the standard semi-empirical
Jeffery-Austin equation of state\cite{JeffAustinEOS} (comparison in
FIG.~\ref{fig:eosFit}(b)), especially considering the fact that the latter fit employs
more than twice as many ($\sim 9$) adjustable parameters.  Beyond
providing a reasonable fit to the equation of state, the key advantage of the present work
is that these results stem directly from a model microscopic Hamiltonian,
which we exploit below to construct a theory for the inhomogeneous fluid.

\begin{figure}
\includegraphics[width=3.375in]{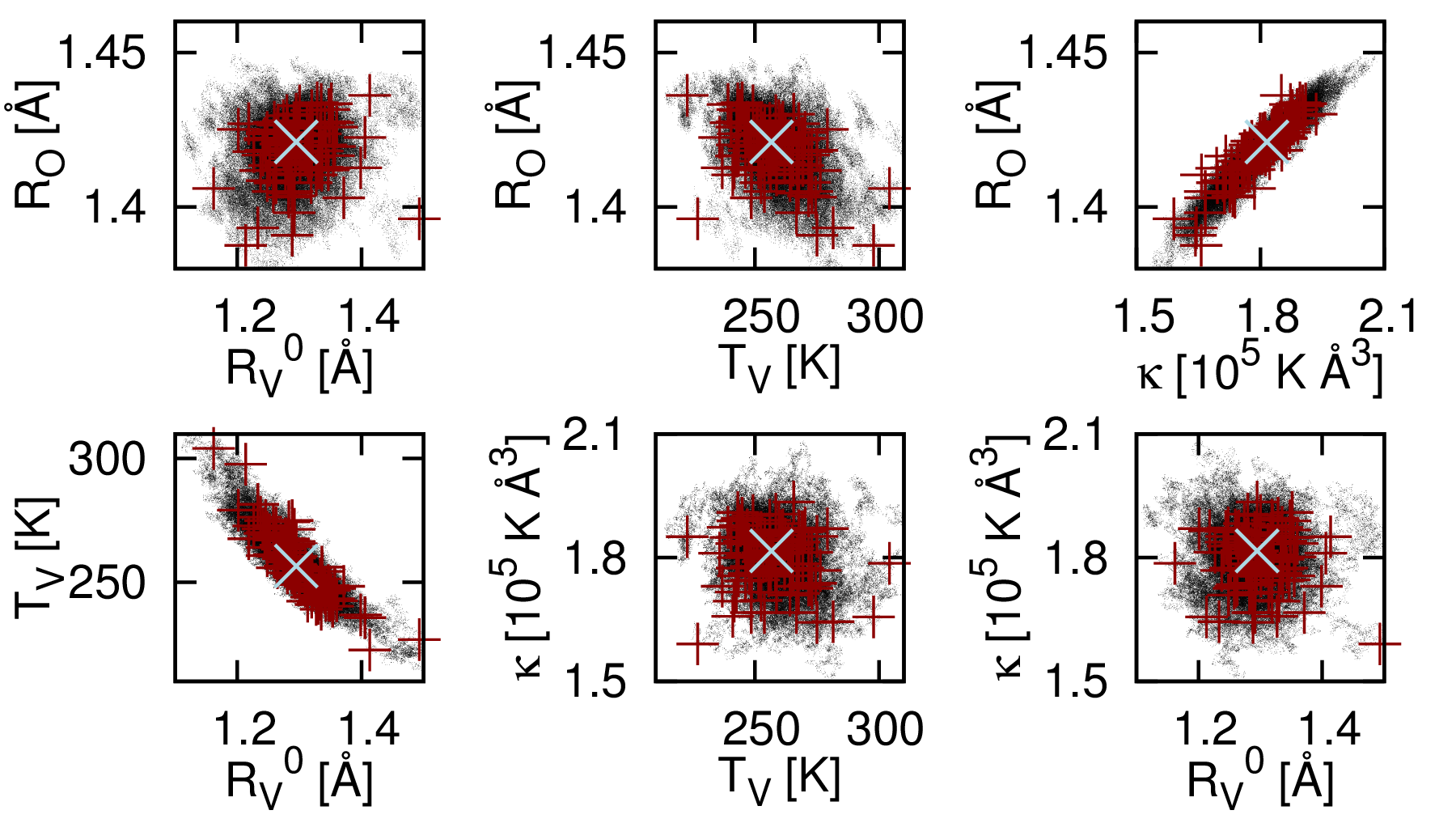}
\caption{Metropolis sampling of a canonical ensemble of parameters,
shown in all six projections of the four-dimensional parameter space.
One hundred random samples ($+$) are drawn from the full set ($\cdot$)
for error estimation of all subsequent results;
$\times$'s mark the optimum parameter set. \label{fig:paramEnsemble}}
\end{figure}

To ensure that our model parameters are indeed independent and
physically meaningful, we employ Bayesian error estimation
following Ref.\citenum{SloppyModels},\citenum{BayesianDFT}.
Specifically, we generate a canonical ensemble of parameter sets
(FIG.~\ref{fig:paramEnsemble}) with a Metropolis walk in parameter space,
where the residual is the `energy' and the `temperature' is $2C_0/N_p$
where $C_0$ is the minimum residual and $N_p=4$ is the number of fit parameters.
The optimum parameters with standard deviations thus estimated are
\begin{flalign}
R_V^{(0)} &= (1.290 \pm 0.049)\mbox{\AA}\nonumber\\
T_V &= (258.7 \pm 12.3) K \nonumber\\
\kappa &= (1.805 \pm 0.074) \times 10^5 K \mbox{\AA}^3 \nonumber\\
R_O &= (1.419 \pm 0.010)\mbox{\AA}. \label{eqn:params}
\end{flalign}

The modest eccentricities of the ensemble slices in FIG.~\ref{fig:paramEnsemble}
indicate that the covariances of these parameters are nominal,
suggesting that the parameters have physical meaning rather than
merely controlling a flexible fit function for the equation of state.

An advantage of this ensemble-of-models approach, which we exploit below,
is that one can estimate how well the fit to bulk data constrains all subsequent
predictions, including those for the inhomogeneous fluid, by evaluating
those predictions for a sampling of the ensemble of parameters
(indicated in FIG.~\ref{fig:paramEnsemble} as +'s),
instead of just the one optimum parameter set.

\section{Model for inhomogeneous liquid} \label{sec:InhomogeneousFluid}
Capturing the behavior of the inhomogeneous fluid requires information beyond
merely the integrated strength $\kappa$ of the pair-potential interaction $U_a(r)$.
This work demonstrates that the simplest next step, including information
about the range of the interaction, suffices to capture surprisingly
well the main features of the short-range correlations in the liquid.
To this end, we employ the attractive-part of the Lennard-Jones potential
\begin{equation}
U_a(r) = \frac{-9\kappa}{8\pi\sqrt{2}\sigma_U^3}
	\left\{ \begin{array}{lc}
		(\sigma_U/r)^6 - (\sigma_U/r)^{12}, & r \ge 2^{1/6}\sigma_U \\
		1/4, & r < 2^{1/6}\sigma_U
	\end{array}\right.
\end{equation}
which has the correct long range $r^{-6}$ tail for the
orientation-averaged interaction of a dipolar fluid.
We fit the range $\sigma_U$ to reproduce the bulk surface tension
at $298$~K (based on calculations with the free-energy functional below),
finding $\sigma_U = 2.62$~\AA. We thereby introduce only one
additional fit parameter in going to the inhomogeneous fluid.

\begin{figure}
\includegraphics[width=3.375in]{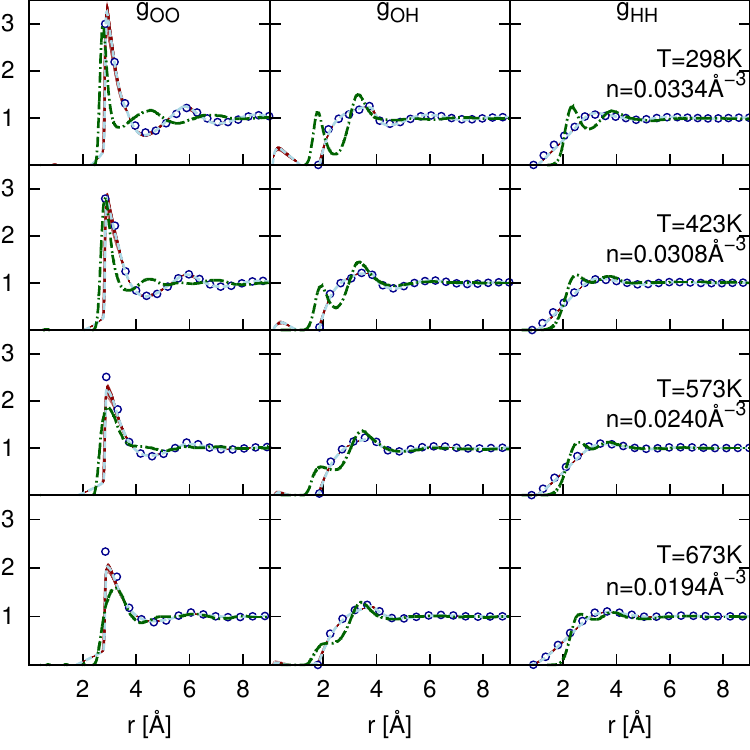}
\caption{Partial radial distributions from the density functional
(\ref{eqn:Functional}) (bundle of thin red lines for the ensemble,
with results for optimum parameters highlighted by dashed line)
and Monte Carlo simulations for the model Hamiltonian (blue circles),
compared to experimental pair correlations of water from
Soper et al.\cite{SoperEPSR} (green dot-dashed line) \label{fig:gXX}}
\end{figure}

To evaluate the viability of this simple model Hamiltonian for
describing the inhomogeneous fluid, we compute its pair correlation functions
(for each of the state points for which experimental correlations
were measured in Soper et al.\cite{SoperEPSR}) directly with
canonical-ensemble Monte Carlo simulations of 2048 molecules.
The comparison (FIG.~\ref{fig:gXX}) between the behavior of this model
microscopic Hamiltonian (circles) and the actual experimental correlations
in {\em physical} water (green dash-dot line) is remarkable given the highly simplified form
for the model.  Although the secondary peaks in the $O$-$O$ correlations of
this model Hamiltonian do appear more at the characteristic distances for a
hard-sphere rather than at those for a tetrahedrally bonded fluid,
the temperature and pressure dependence of the locations and heights
of the first $O$-$O$ peak compare reasonably to water.
Similarly, although the first two peaks of the $O$-$H$ and $H$-$H$
correlations are fused into a single broader peak, the general location
and particle content of these peaks are in reasonable agreement
for such a simple model.  These details could be corrected in future work
by fitting perturbation pair potentials of zero integral
$\Delta U_{\alpha\beta}(r)$ with $\alpha,\beta\in\{O,H\}$
to the experimental correlation data, but the focus of the present work
is the quality of predictions which can be made from a simple
microscopic model with very few adjustable parameters (five)
constrained purely by the macroscopic data.

Having established a \emph{short-ranged} microscopic model Hamiltonian which reproduces
relatively well the experimental correlations in water, we turn next to development
of a corresponding free-energy functional. The form of this functional,
\begin{flalign}
\Phi[\psi] = \Phi_{\textrm{id}} + \Phi^{\textrm{ex}}_{\textrm{HS}}
	+ \Phi_{b} + \frac{1}{2}\int n_O \left(U_a\ast n_O\right),
\label{eqn:Functional}
\end{flalign}
mirrors the equation of state (\ref{eqn:EOS}), and is composed
of the ideal gas free energy, hard sphere excess functional,
bonding correction and mean field perturbation.

We start with the exact grand free energy functional for the
ideal gas of rigid molecules and thereby restore exact treatment
of the intramolecular bond-angle constraints; this approach is consistent
with Wertheim theory since the latter yields the exact rigid-molecular
ideal gas pressure $p_{id}=nT$ at $\mathcal{O}(n)$ in the uniform limit.
The free energy of the inhomogeneous ideal gas with chemical potential $\mu$
in external site potentials $V_\alpha(\vec{r})$ is written as
\begin{flalign}
\Phi_{\textrm{id}}[\psi,n[\psi]] =& \sum_\alpha \int d\vec{r}
		n_\alpha(\vec{r})(V_\alpha(\vec{r})-\psi_\alpha(\vec{r})) \nonumber\\
	&- (\mu+T)\int d\vec{r} n_O(\vec{r}),
\end{flalign}
employing ideal gas effective potentials \cite{RISM1,RISM2} $\psi_\alpha(\vec{r})$
for $\alpha\in\{O,H,V\}$ as the sole independent variables.\cite{LischnerHCl}
Here, the site densities are dependent variables computed using
\begin{flalign}
n_\alpha(\vec{r})=\frac{\delta}{\delta\psi_\alpha(\vec{r})}
	\int \frac{d\omega d\vec{r}'}{4\pi^2} \exp \frac{-1}{T}\sum_{\alpha',i}
		\psi_{\alpha'}(\vec{r}'+\omega\circ \vec{R}_{\alpha' i}),
\label{eqn:NfromPsi}
\end{flalign}
where $\omega\in SO(3)/\mathbb{Z}_2$, where $\omega\circ$ denotes the
corresponding rotation for a vector, and where $\vec{R}_{\alpha i}$
are the site coordinates for a molecule in the reference orientation
centered at the origin with $i=1$ for $\alpha=O$ and $i\in\{1,2\}$
for $\alpha\in\{H,V\}$. Note that we have simplified the above
expression using the $\mathbb{Z}_2$ rotation symmetry of the
molecule about its dipole axis.

To treat the hard sphere mixture excess free energy
$\Phi^{\textrm{ex}}_{\textrm{HS}}$, we use the `White-Bear mark II'
version of fundamental measure theory\cite{WhiteBearFMT_MarkII}
(incorporating Tarazona's tensor modifications\cite{TarazonaFMT})
\begin{flalign}
\Phi^{\textrm{ex}}_{\textrm{HS}} &= T \int \left[\begin{array}{c}
		n_0 \log\frac{1}{1-n_3} + \frac{n_1n_2 - \vec{n}_{v1}\cdot\vec{n}_{v2}}{1-n_3} f_2(n_3) + \\
		\frac{n_2^3 - 3n_2\vec{n}_{v2}^2
			+ 9\vec{n}_{v2}\bar{n}_{m2}\vec{n}_{v2}-\frac{9}{2}\textrm{Tr}\bar{n}_{m2}^3
			}{24\pi(1-n_3)^2} f_3(n_3)
	\end{array}\right], \nonumber\\
\end{flalign}
in terms of the scalar, vector and tensor weighted densities
$n_i = w_i^O \ast n_O + w_i^V \ast n_V$ for $i\in\{0,1,2,3,v1,v2,m2\}$,
where
\begin{equation*}
	\begin{array}{rl}
		f_2(n_3) &= 1 + \frac{n_3(2-n_3) + 2(1-n_3)\log(1-n_3)}{3n_3}\\
	\textrm{and }
		f_3(n_3) &= 1 - \frac{2 n_3 - 3 n_3^2 + 2 n_3^3 + 2(1-n_3)^2\log(1-n_3)}{3 n_3^2}.
	\end{array}
\end{equation*}
(See the comprehensive review by R. Roth\cite{FMTreview} for details.)
Note that this functional corresponds exactly to the hard sphere
excess pressure (\ref{eqn:pExHS}) in the uniform fluid limit.

Next, $\Phi_{b}$ accounts for the tangential bonding constraints
on the hard-sphere exclusion effects; note that the contribution
from Wertheim pertubation to the ideal gas part has been absorbed
into the exact rigid-molecule ideal gas free energy $\Phi_{\textrm{id}}$.
The Helmholtz-energy density for this term in the uniform fluid limit
is determined from the third term of (\ref{eqn:EOS}) to be
$-2nT\log g^{\textrm{HS}}_{OV}(\sigma_{OV})$, which we generalize
to the inhomogeneous version
\begin{flalign}
\Phi_{b} = \int \frac{-2n_0T}{3}\log
	\left[ \frac{1}{1-n_3} + \frac{\zeta n_2 R_{\textrm{hm}}}{(1-n_3)^2}
		+ \frac{2\zeta n_2^2 R_{\textrm{hm}}^2}{9(1-n_3)^3} \right],
\end{flalign}
with the vector correction factor $\zeta=1-|\vec{n}_{v2}|^2/n_2^2$.
We include this factor here following the spirit of Yu et al.,\cite{WuBonding}
where $\zeta$ was introduced in analogy with the occurence of the vector 
weighted densities in the hard sphere mixture functional
in order to improve agreement with Monte Carlo calculations.
Finally, the last term in (\ref{eqn:Functional}) describes the
attractive perturbation potential within a mean-field picture.

The partial radial distribution functions implied by the free energy functional
(\ref{eqn:Functional}), as evaluated from its analytic second variational
derivatives using the Ornstein-Zernike relation, are in excellent agreement
with the Monte Carlo simulations (circles and corresponding curve in FIG.~\ref{fig:gXX}).
(The minor artifacts in the interior of the hard cores are caused by the
bonding correction, whose inhomogeneous generalization is not perfect.)
The small spread in these results with variation of parameters
in the ensemble exemplifies how tightly the bulk data
indeed constrain these predictions within the assumed model.

\section{Predictions for the inhomogeneous liquid}
To evaluate the predictions of the above density functional for the
inhomogeneous fluid, we perform direct minimization of (\ref{eqn:Functional})
using the nonlinear conjugate gradients algorithm \cite{NCG} with the values
of $\psi_\alpha(\vec{r})$ on a discretized grid as the independent variables.
The orientation integrals involved in evaluating the site densities
from the site potentials (\ref{eqn:NfromPsi}) are discretized
using quadratures on $SO(3)/\mathbb{Z}_2$. The calculations
presented below are performed on radial or planar $d=1$ dimensional grids
\cite{Note2}
%
where the azimuthal symmetry simplifies the orientation quadrature
from $SO(3)\equiv\mathbb{S}_2\times\mathbb{S}_1$ to $\mathbb{S}_2$,
which we tesselate using a recursively subdivided icosahedron.
\cite{Note1}

\begin{figure}
\includegraphics[width=3.375in]{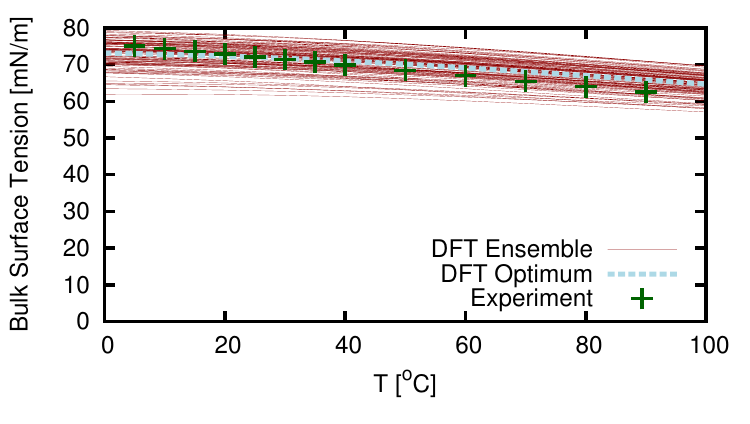}
\caption{Energy of the vapor-liquid interface as a function
of temperature, compared to the experimental values for
surface tension.\cite{LangeHandbook}
\label{fig:sigmavsT}}
\end{figure}

\begin{figure}
\includegraphics[width=3in]{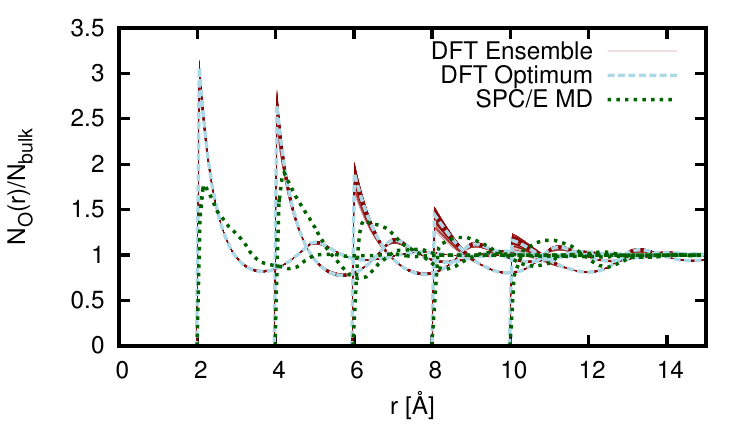} \\
\includegraphics[width=3in]{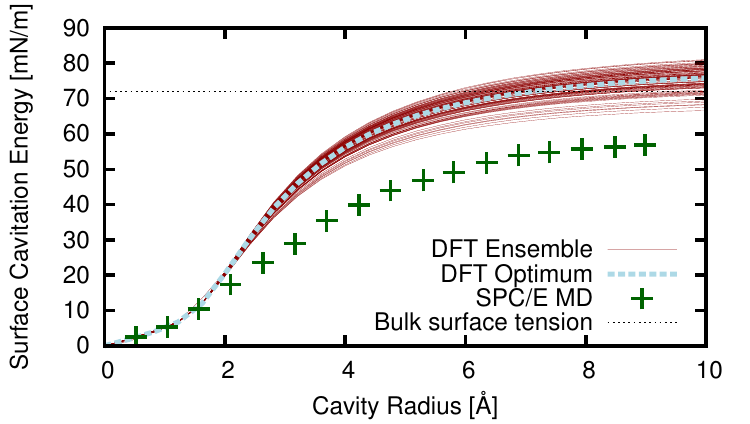}
\caption{Comparison of density functional predictions
with SPC/E molecular dynamics results\cite{HardSphereSPCE} for
(a) Radial distribution around spheres of radii
2, 4, 6, 8, and 10\AA\ that exclude the oxygen sites and
(b) the variation of solvation energy of such spheres
per surface area with radius. \label{fig:hardsphere}}
\end{figure}

We find remarkable agreement with available data for the behavior and
free energies of inhomogeneous aqueous systems, especially given that
{\em only} bulk data, including surface tension at a {\em single}
temperature, were employed in determining the limited number of
parameters in the functional.  For example, FIG.~\ref{fig:sigmavsT}
compares our prediction of the temperature dependence of the
interfacial energies with experimental data.  Over the entire
range of accessible temperatures at ambient pressure, we find the
experiment to lie within the relatively narrow variations within our
ensemble of models.  Moving beyond planar interfaces,
FIG.~\ref{fig:hardsphere} explores the radial distribution around hard
spheres and the variation of free energy of hard-sphere insertion with
radius, and demonstrates that the predictions of our model are in
qualitative agreement with the SPC/E molecular dynamics results.
\cite{HardSphereSPCE}  The contact densities and the free energies
from our model are somewhat higher than those from the SPC/E model
results, a situation which could potentially be improved in future work
by including additional perturbation pair-potentials.

\begin{figure}
\includegraphics[width=3.375in]{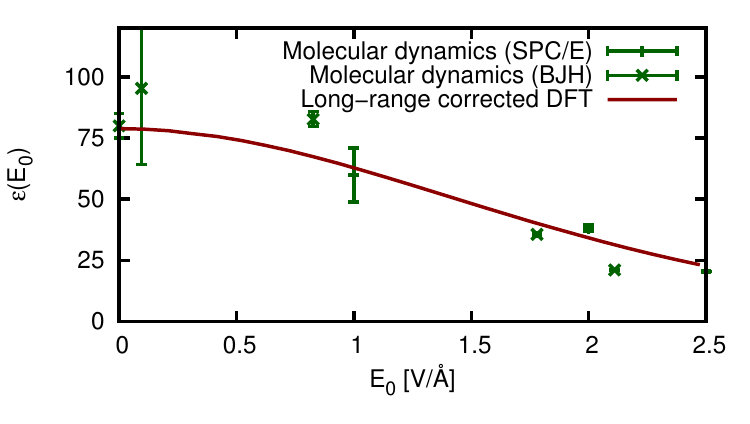}
\caption{Nonlinear dielectric response: variation of relative
dielectric constant $\epsilon$ with externally applied field $E_0$
at ambient conditions in comparison to SPC/E\cite{NonlinearEpsSPCE}
and BJH\cite{NonlinearEpsBJH} molecular dynamics simulations.
By construction, the linear dielectric constant matches experiment
due to the long-range corrections.\label{fig:nonlineareps}}
\end{figure}

In addition to the bulk and short-ranged correlations described above,
a successful theory of solvation requires accurate dielectric response.
Following Lischner et al.,\cite{LischnerH2O} we add a scaled mean-field
long range electrostatic correction
\begin{flalign}
\Phi_\epsilon = \frac{A_\epsilon(T)}{2}\sum_{\alpha,\beta\in\{O,H\}}
	Z_\alpha Z_\beta \int n_\alpha K\ast n_\beta ,
\end{flalign}
where the site charges $Z_\alpha$ are taken to be the SPC/E values
\cite{SPCE} and $K\equiv \frac{4\pi}{G^2(1+(G/G_c)^4)}$ with $G_c=0.33$
is the Coulomb kernel with a high frequency (short range in space) cutoff \cite{LischnerH2O}.
The prefactor $A_\epsilon(T)=1 - T/(7.35\times10^3~\textrm{K})$ serves to correct
for dipole correlations beyond mean field, and is fit to reproduce
the bulk dielectric constant {\em at small field}. Fig.~\ref{fig:nonlineareps}
shows that the nonlinear response at high fields (which is not fit)
is well captured by the interplay between $\Phi_\epsilon$ and $\Phi_{\textrm{id}}$.

{\em Conclusion ---} We have constructed a computationally tractable
free-energy functional for studies of inhomogeneous water based upon a
microscopic Hamiltonian constrained by experimental data for the bulk
equation of state.  Following this approach gives a remarkably
high-quality fit to the equation of state with only four tightly
constrained parameters.  With one additional parameter, the range of
the model interaction, the resulting functional captures the free
energies associated with inhomogeneous systems such as the
liquid-vapor interface and the embedding free energy of microscopic
objects, as well as essential features of the partial radial
distributions and density profiles around microscopic objects. With
long-range corrections, the model gives an accurate description of the
non-linear dielectric response.  The model thus shows good promise for
capturing the key quanties which require description in solvation
studies.  In future work, further details may be captured with
suitable perturbation of the pair potentials constituting the
underlying microscopic Hamiltonian.

This work was supported as a part of the Energy Materials Center at Cornell (EMC$^2$),
an Energy Frontier Research Center funded by the U.S. Department of Energy,
Office of Science, Office of Basic Energy Sciences under Award Number DE-SC0001086.
We would like to thank N. W. Ashcroft for valuable suggestions regarding the
microscopic Hamiltonian and J. P. Sethna for recommending the Bayesian analysis.

%

\end{document}